\documentclass[aps]{revtex4}
\usepackage{axodraw}
\usepackage{graphicx}
\usepackage{graphics}

\newcommand{\ra}{\rightarrow}
\newcommand{\bi}{\bibitem}

\begin{document}
\bibliographystyle{revtex}

\preprint{IFT-2001/40}
{\begin{flushright}
IFT-2001/40 \\ [1.5ex]
\end{flushright}}
\title{The $(g-2)_{\mu}$ data and the lightest Higgs boson in 2HDM(II)}


\author{Maria Krawczyk}
\affiliation{Institute Theoretical Physics, Warsaw University, ul. Hoza 69, 
00-681 Warsaw, Poland}

\begin{abstract}
The present limits on the lightest Higgs boson in 2HDM (II)
in light of the new E821 measurement of $g-2$ for the muon
 are discussed.
\end{abstract}
\pacs{13}
\maketitle

The precision measurement of $g-2$ for the muon is expected to shed 
light on  "new physics". 
Here \cite{Krawczyk:2001pe}  we discuss constraints  on  
the lightest neutral Higgs boson in 2HDM (II) which can be drived 
from the new E821 measurement  based on 1999 data
\cite{Brown:2001mg}.
 A current mean of experimental 
results  for $(g-2)_{\mu}$ is (from \cite{Brown:2001mg})
$$a_{\mu}^{exp}\equiv{{(g-2)_{\mu}^{exp}}/{2}}=11~659~203~(15)
\cdot 10^{-10},$$
where the  accuracy of this result (in parentheses) approaches the 
size of electro-weak contribution, 
$a_{\mu}^{EW}$.
The ultimate accuracy  of the E821 experiment is    4 $\cdot 10^{-10}$.
The Standard Model prediction for $a_{\mu}$ 
consists of the QED, hadronic and EW contributions:\\
\centerline{$a_{\mu}^{SM}=a_{\mu}^{QED}+a_{\mu}^{had}+a_{\mu}^{EW}.$}
Both the  QED and EW contributions to $a_{\mu}^{SM}$ are well under control.
The   predictions for the hadronic contribution, $a_{\mu}^{had}$,
differ considerably among themselves.
Its uncertainty is presently  of order $ (7 - 10)  \cdot 10 ^{-10}$,
with the dominant  error coming  from 
  the leading vacuum polarization contribution (see  
 discussions in \cite{Czarnecki:2001pv},  also     
\cite{Yndurain:2001qw}).
A new controversy  of the 
light-by-light scattering contribution (a sign !) has appeared recently 
\cite{Knecht:2001qf}.

The difference between the experimental data, $a_{\mu}^{exp}$,
 and the Standard Model (SM) 
prediction, $a_{\mu}^{SM}$, defines the room for  "new physics".
Obviously the uncertainties of the hadronic contributions  influence  
 the estimation of a size of  new effects.
To illustrate  the present situation  we  calculate  
   95\% CL  intervals ($lim (95\%)$) for an 
allowed new contribution, $\delta a_{\mu}$,  
using  two representative SM predictions \cite{Czarnecki:2001pv}:  
one based on  the  calculation of leading  vacuum polarization diagrams 
by Davier and H\"ocker (DH)\cite{Davier:1998si} 
and the  other by   Jegerlehner (J2000)\cite{Eidelman:1995ny},
with a smaller and larger 
$a_{\mu}^{had}$ (and its uncertainty), respectively.
 
The obtained intervals we apply  to constrain the parameters of the 
non-supersymmetric, CP conserving  2HDM ('Model II') \cite{hunter}. 
This model,   based on the two doublets of 
complex scalar fields,  predicts existence of five Higgs particles: 
two neutral Higgs scalars $h$ and $H$, one neutral pseudoscalar $A$, 
and a pair of charged Higgses $H^{\pm}$. 
In this  model  {\sl one} light neutral 
Higgs boson, $h$ or $A$,  with mass even below 20 GeV is still allowed by
 the  LEP \cite{Krawczyk:1999kk,Yu}
and low energy experiments  \cite{keh,pich}  in contrast to
 the  SM Higgs boson which should be heavier  than 114.1
 GeV (95 \% CL) \cite{MSSM}.
 Previous (CERN)  data  and the SM prediction(s) for $a_{\mu}$
were used by us to derive the 
constraints (based on the one-loop calculation \cite{old})  
 for $h$ or $A$ in 2HDM (II) \cite{Krawczyk:1997sm}. 
A very small improvement with comparison to the LEP limits were obtained.
In this paper we use new E821 data and apply the  two-loop approach 
\cite{Chang:2001ii} to derive tight constraints on the Yukawa couplings to
 muon of $h$ and $A$  (similar results  are in  \cite{Chang:2001ii}, 
see also \cite{Dedes:2001nx,Krawczyk:2001pe} 
for one-loop results). Next we combine these constraints with other
limits to derive  current constraints for the lightest Higgs boson in 
2HDM(II).

 We study separately  case A  
and  case B based on the DH and J2000 
calculation of the leading 
vacuum polarization diagram, respectively. 
The contributions due to higher order hadronic corrections 
 and the light-on-light scattering
 are taken from \cite{Kinoshita:1985it} 
and \cite{ll}, respectively.    
In the table below we collect, following \cite{Czarnecki:2001pv}, 
the corresponding SM contributions (and their uncertainties).
From   $\Delta a_{\mu}$, the difference of the central 
values $a_{\mu}^{exp}-a_{\mu}^{SM}\equiv \Delta  a_{\mu}$, and 
the error for this quantity, $\sigma$,  one can  calculate an allowed
 at chosen confidence level (CL) interval of an additional contribution. 
We estimate  $\sigma$ by adding in 
quadrature 
$\sigma_{exp}$ and $\sigma_{tot}$.   
 Assuming a Gaussian distribution   
  we    calculate in both cases, A and B, 
the  allowed at 95\% CL $\delta a_{\mu}$ regions, symmetric  
around $\Delta a_{\mu}$ (see table below).
$$
\begin{array}{lrr}
case  &~{ \rm {A}}~[{ \rm in}~ 10^{-11}]&~{\rm {B~[in}}~ 10^{-11}] \\  
\hline
{\rm QED}   &~~~116~584~706 ~~~~(3)
  &~~~116~584~706~ ~~~~(3) \\
{\rm had}   & 6~739~~(67) &  6~803 ~~(114)  \\
{\rm EW}    & 152 ~~~~(4)  &  152 ~~~~~(4)   \\
\hline
{\rm tot}   &116~59 1 ~597 ~~(67)   & 116 ~591 ~661 ~~(114)\\
\hline\hline 
\Delta  a_{\mu}(\sigma)    &426 (165) &~362 (189)  \\
\hline
{ lim(95\%)} &102\le\delta a_{\mu} \le 750 
     &~-8.65\le\delta a_{\mu} \le 733 \\      
\hline 
\end{array}   
$$

We see that 
at  95 \%CL
the more conservative estimation of the hadronic contribution to 
$a_{\mu}^{SM}$ (case B) leads to  both  the negative and positive 
 $\delta a_{\mu}$, while in case A $\delta a_{\mu}$ is  of
 a positive sign only.
As a consequence, the 95 \%CL interval
leads in case  A to  an {\sl allowed   positive} contribution 
({\sl an allowed  band})
 and at the same time to  the {\sl exclusion} of the negative contribution. 
For the case B,  the positive  (negative) contribution  is only  
bounded from above (below) ({\sl   upper limits} for the absolute 
value of the new contribution).
This reflects the fact   that the  SM prediction lies within 
the 95\% CL interval 
for case B, while for case A it is outside the corresponding interval.\\
\vspace*{0.3cm}
\begin{center} 
\begin{picture}(300,80)(0,0)
\Text(30,90)[r]{$\gamma$}
\Text(45,30)[r]{h,A}
\Text(0,5)[r]{$\mu$}
\Text(82,5)[r]{$\mu$}
\Photon(40,90)(40,70){3}{3}
\ArrowLine(0,0)(40,70)
\ArrowLine(40,70)(70,0)
\DashLine(25,40)(50,40){5}
\end{picture} \\

\vspace*{0.2cm}
\begin{picture}(200,100)(-150,-110)
\Text(30,90)[r]{$\gamma$}
\Text(80,30)[r]{$\gamma$}
\Text(10,30)[r]{h,A}
\Text(-3,5)[r]{$\mu$}
\Text(90,5)[r]{$\mu$}
\Text(65,60)[l]{$f(W,H^+)$}
\Photon(40,90)(40,70){3}{3}
\Photon(60,40)(70,20){3}{3}
\ArrowLine(0,0)(10,20)
\ArrowLine(10,20)(70,20)
\ArrowLine(70,20)(80,0)
\ArrowArc(40,50)(20,2,1)
\DashLine(10,20)(20,40){5}
\end{picture} \\ 
\vskip -3.7cm
{ One- and two-loop diagrams. The $W^+$ and $H^+$
loops contribute only for a $h$ exchange.}
\end{center}

We apply the obtained intervals  $\delta a_{\mu}$
 to  constrain  parameters of the 2HDM (II) using a simple approach,
where only one neutral Higgs boson, $h$ or $A$, contributes.
In the one-loop calculation (based on the one-loop diagram, see above) 
a light scalar scenario leads to the positive, whereas the
one with a light pseudoscalar to the negative contribution to $a_{\mu}$, 
independently of mass.
In the two-loop analysis, based on a sum of the one- and two-loop
 (fermionic and bosonic) diagram  contributions, 
 the situation changes drastically.
Now the positive (negative) contribution can be ascribed
to a scalar $h$ with mass below (above) 5 GeV or  a  pseudoscalar $A$ with
mass above (below) 3 GeV. For $h$ we assumed $\chi_V^h$=0, so
the $W$-loop does not contribute. 
A $H^+$ loop, for $M_{H^+}$ = 400, 800 GeV and infinity 
(and with parameter $\mu=0$ in the $hH^+H^-$ coupling), is included in 
the analysis.

\begin{figure}[ht]
\begin{center}
\hspace*{-3.7cm}
\includegraphics{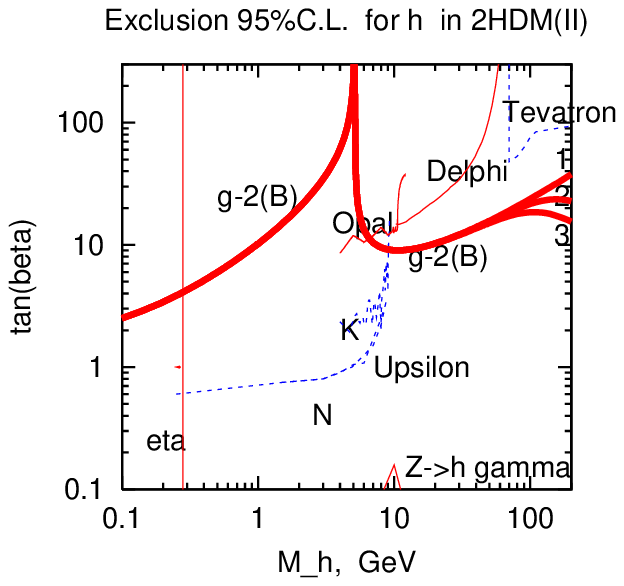}%
\includegraphics{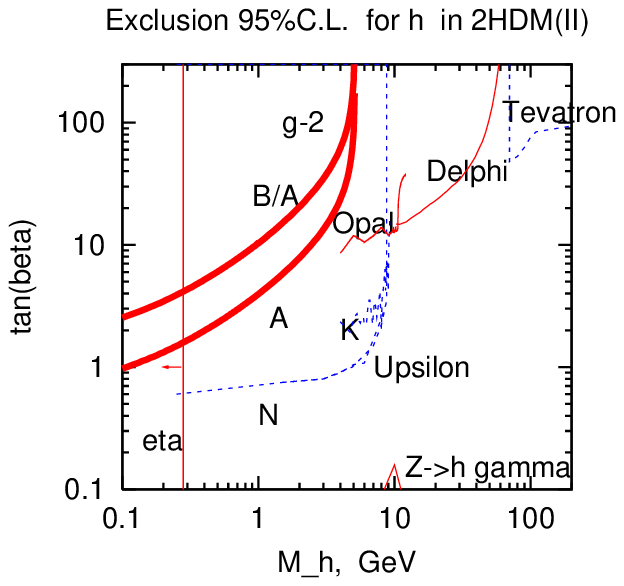}%
\end{center}
\caption{ The  95\% CL upper and lower limits  for the Yukawa coupling 
$\chi_d^h$ for a scalar $h$ ($\chi_d^h=\tan \beta$ for  vanishing 
coupling to gauge bosons $\chi_V^h=0$)  as a function of mass. 
Left: The tick lines  ``$g-2(B)$''  
give upper limits in case B. Line 1 (2,3) is obtained  for  mass of $H^+$ 
 equal to infinity (800, 400 GeV). Right: Similar results for case A, 
allowed region between upper tick line B/A (which
 coincides with case B) and lower tick line A.}
\label{fig:P3-43fig1}
\vspace*{-4.8cm}
\begin{center}
\hspace*{-3.7cm}
\includegraphics{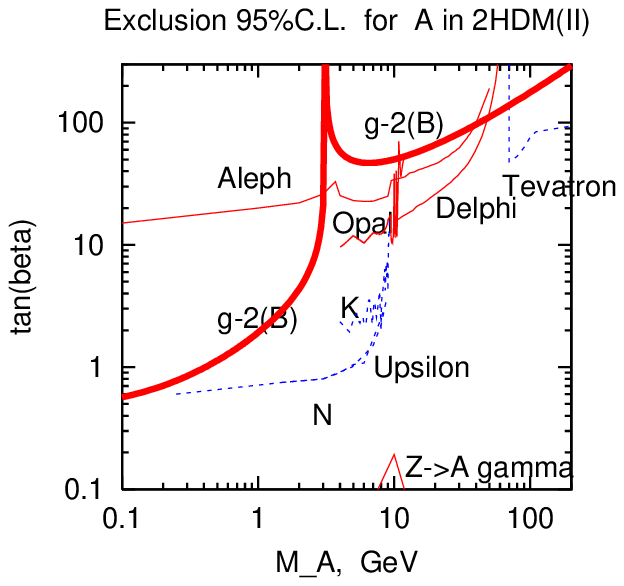}%
\includegraphics{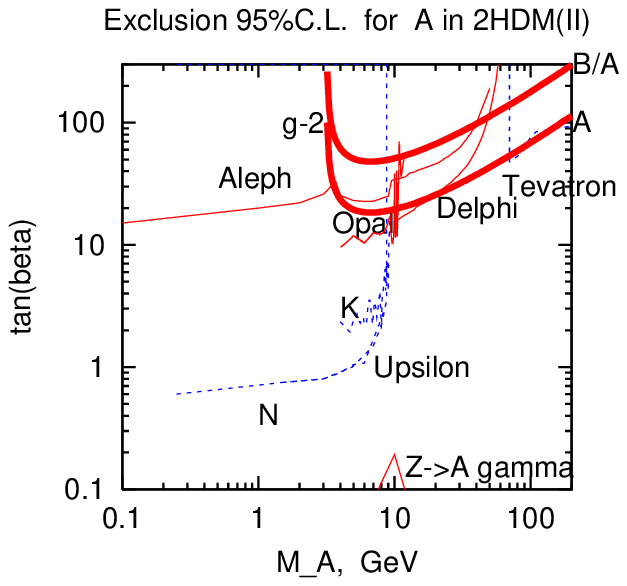}%
\end{center}
\caption{As in Fig.1 for a pseudoscalar $A$ ($\chi_d^A = \tan \beta $).}
\label{fig:P3-43fig2}
\end{figure}

Our results are presented in Fig.\ref{fig:P3-43fig1} and Fig.
\ref{fig:P3-43fig2} 
(tick lines) together with current upper limits     
from the Yukawa process (ALEPH, DELPHI and OPAL results)
and lower limits from the $Z\ra h(A) \gamma$ \cite{Krawczyk:1999kk}. 
In addition the upper 90\% CL limits  from 
 the  $\Upsilon$ decay (K,N), rescaled by a factor 2, and from the 
TEVATRON \cite{TEV} are presented. 
Constraints based on  case B  lead to an improvement
of the existing  upper limits for  $h$  for mass above 10 GeV
(Fig.\ref{fig:P3-43fig1}(left))  and for $A$ above 50 GeV (Fig.
\ref{fig:P3-43fig2}(left)). For 
 case A  our results are in form of allowed regions
for mass below 5 GeV for $h$ (Fig.\ref{fig:P3-43fig1}(right)) 
and for mass above 3 GeV for $A$
(Fig.\ref{fig:P3-43fig2}(right)) (see also \cite{Chang:2001ii}). 

This two-loop analysis based on new $(g-2)_{\mu}$ data and on the 
 DH estimation of  $a_{\mu}^{had}$ (case A)
 if combined with   constraints  from other experiments
allows in  the 2HDM (II) for an existance  of a pseudoscalar 
with mass between $\sim$ 25 GeV and 70 GeV,
 and $\tan \beta$ above 30. A light scalar is excluded.
This latter result  is in agreement with a conclusion of theoretical analysis
 \cite{Kanemura:1999xf}.
\begin{acknowledgments}
I thank P. Zalewski, M. Kobel and F. Akesson for sending  their new results
on Yukawa process.
I am grateful organizers of this excellent meeting for the financial support.
{Supported in part
 by Polish Committee for Scientific Research,
 Grant5 P03B 121 20,  and the EC 5th framework 
contract HPRN-CT-2000-00149.}
\end{acknowledgments}

\end{document}